\documentclass[12pt,letterpaper]{article}
\usepackage{amsmath,amssymb,calc}
\usepackage{graphicx}

\newcommand\be{\begin{equation}}
\newcommand\ee{\end{equation}}
\newcommand{\bea}{\begin{eqnarray}}
\newcommand{\eea}{\end{eqnarray}}

\newcommand{\la}{\langle}
\newcommand{\ra}{\rangle}

\newcommand{\nn}{\nonumber}
\newcommand{\pd}{\partial}

\def\id{\protect{{1 \kern-.28em {\rm l}}}}

\def\unit{\relax{\rm 1\kern-.26em I}}

\def\id{\protect{{1 \kern-.28em {\rm l}}}}
\begin{document}

\begin{titlepage}
\begin{center}
\hfill \\
%\hfill {\tt ......}\\
%\vskip 15mm
\vspace{2cm}
{\Large {\bf Flux Vacua Attractors in Type II on \\ $SU(3)\times SU(3)$ Structure\\[3mm] }}

\vskip 10mm

{\bf Lilia Anguelova}

\vskip 4mm
{\em Department of Physics}\\
{\em University of Cincinnati, Cincinnati OH 45221, USA}\\
{\tt anguella@ucmail.uc.edu}

\vskip 6mm

\end{center}

\vskip .1in
\vspace{2cm}

\begin{center} {\bf Abstract }\end{center}

\begin{quotation}\noindent
We summarize and extend our work on flux vacua attractors in generalized compactifications. After reviewing the attractor equations for the heterotic string on $SU(3)$ structure manifolds, we study attractors for $N=1$ vacua in type IIA/B on $SU(3)\times SU(3)$ structure spaces. In the case of vanishing RR flux, we find attractor equations that encode Minkowski vacua only (and which correct a previous normalization error). In addition to our previous considerations, here we also discuss the case of nonzero RR flux and the possibility of attractors for AdS vacua.
\end{quotation}
%%%%%%%%%%%%%%
\vfill
%%%%%%%%%%%%%%%%%%
%\flushleft{\today}
%%%%%%%%%%%%%%%%%%

\end{titlepage}

\eject

%\tableofcontents

\section{Introduction}

Background fluxes are a necessary ingredient in order to achieve moduli stabilization in string compactifications \cite{DRS, GKP}. Naturally, those fluxes backreact on the geometry and thus lead to more involved internal manifolds than the familiar CY 3-folds. It has been known for awhile that supersymmetry requires the internal space of a generic heterotic flux compactification to be a manifold with $SU(3)$ structure \cite{AS}\footnote{See \cite{Het} for the reformulation of Strominger's result in modern $SU(3)$ structure language.}. Recently, it was found that for type IIA/B the analogous requirement is for the internal space to have $SU(3)\times SU(3)$ structure \cite{GLW}. In fact, the low energy effective theory of such type II compactifications has $N=2$ supersymmetry. However, the $N=1$ vacua obtained in this context encompass the most general IIA/B flux vacua with $N=1$ supersymmetry \cite{GMPT}, in the realm of geometric compactifications.

Manifolds with $SU(3)\times SU(3)$ structure are, in principle, much less understood than CY(3)s. The most suitable framework for their study is generalized complex geometry \cite{NH, MG}. The latter deals with objects defined on the tangent plus cotangent bundle of a manifold, $T\oplus T^*$, instead of just on $T$. This, in particular, allows a unified description of complex and symplectic geometry. In this framework a generalized almost complex structure ${\cal I}$ is a map from $T\oplus T^*$ to itself, which squares to $-1$. The integrability condition is that the $+i$ eigen-bundle of ${\cal I}$ be closed under the Courant bracket.\footnote{Note that this is a natural generalization of the integrability condition of an ordinary almost complex structure, which is the closedness of its $+i$ eigen-bundle under the Lie bracket.} Each such ${\cal I}$ corresponds to a pure $SO(6,6)$ spinor $\Phi$ (here we assume that ${\rm dim} T = 6$, as is the case in string compactifications). Spaces with $SU(3)\times SU(3)$ structure are characterized by a pair of pure spinors $\Phi_+$ and $\Phi_-$, which can be viewed respectively as even and odd elements of $\Lambda^{\bullet} T^*$. In other words, $\Phi_+$ and $\Phi_-$ are sums of, respectively, even and odd forms of different degrees. They are the generalizations of the familiar K\"{a}hler form $J$ and holomorphic 3-form $\Omega$ that together define a CY(3). The geometric moduli of a string compactification on an $SU(3)\times SU(3)$ structure manifold arise from the deformation spaces of $\Phi_{\pm}$. These deformation spaces have been shown to have special K\"{a}hler geometry \cite{NH, GLW, CB}. Hence, it seems natural to expect that the $N=1$ flux vacua of such generalized compactifications can be encoded in attractor equations similar to the black hole (BH) attractors in supersymmetric field theories, as argued in \cite{RK} for a particular class of type IIB compactifications. In fact, things are not that straightforward at all as the appropriate $N=1$ coordinates are such that one of the two moduli spaces is only K\"{a}hler, but {\it not} special K\"{a}hler.\footnote{In type IIA this is the moduli space of odd forms, whereas in type IIB the one of even forms. We will give more details on this later on.} In \cite{LA}, we investigated attractors in such generalized IIA/B compactifications.

The attractor mechanism was first discovered in the studies of black holes in $N=2$, $d=4$ supergravity coupled to vector multiplets \cite{BHattr}. More precisely, it was shown that the extrema of the relevant black hole scalar potential are given by the solutions of a system of algebraic equations, called attractors. This system determines the values of the BH moduli at the horizon in terms of the BH electric and magnetic charges. The key property that allows the derivation of the BH attractor equations is the special K\"{a}hler geometry of the relevant moduli space. This, together with the similarity between the BH potential and the scalar potential in $N=1$ supergravity, led \cite{RK} to write down attractor equations for $N=1$ flux vacua in type IIB orientifold compactifications that inherit special K\"{a}hler properties from the unorientifolded $N=2$ theory. In this case, the attractor equations determine the values of the compactification moduli at a given vacuum in terms of the corresponding background fluxes. We find attractor equations (that correct a normalization error in \cite{LA}) for a much broader class of $N=1$ type IIA/B flux vacua, namely the most general Minkowski ones obtained by compactifying on $SU(3)\times SU(3)$ structure spaces. In addition to the case of zero RR flux considered in \cite{LA}, here we also discuss the case of non-vanishing RR flux and the possibility for $N=1$ AdS attractors. The conceptual value of the flux vacua attractors is in that they provide a (simpler) reformulation of the problem of minimization of the scalar potential and hence give a new technical tool for the systematic study of moduli stabilization.

\section{Heterotic on $SU(3)$ Structure}

Before turning to the more involved type II on $SU(3)\times SU(3)$ structure case, it is instructive and useful to consider first the heterotic string on $SU(3)$ structure manifolds. The latter are a special case of $SU(3)\times SU(3)$ structure and are characterized by the existence of a two form $J$ and a 3-form $\Omega$, as is a CY(3). However, unlike the $SU(3)$ holonomy case, now generically $dJ\neq 0$ and $d\Omega \neq 0$. For more on $SU(3)$ structure manifolds, see \cite{Het}.

To introduce the geometric and B-field moduli of such a compactification, one expands: 
\be \label{JOexp}
e^{-J_c} = X^{\cal A} (t) \omega_{\cal A} - G_{\cal A} (t) \tilde{\omega}^{\cal A} \, , \qquad
\Omega = X^{\cal I} (z) \alpha_{\cal I} - G_{\cal I} (z) \beta^{\cal I} \, ,
\ee
where $J_c = B + i J$ and $\{ \omega_{\cal A} , \tilde{\omega}^{\cal B} \}$ is a basis for the even forms (i.e. 0-, 2-, 4- and 6-forms), whereas $\{ \alpha_{\cal I} , \beta^{\cal J} \}$ is a basis for the 3-forms.\footnote{Considering $e^{-J_c}$, instead of $J_c$, is useful in view of the $SU(3)\times SU(3)$ case that we will study in the next section.} Also, in (\ref{JOexp}) $t^{\alpha}$ denote the K\"{a}hler moduli and $z^i$ - the complex structure ones. Note that the basis forms satisfy the relations
\bea \label{pairings}
\int \la \alpha_{\cal I} , \beta^{\cal J} \ra = - \int \la \beta^{\cal J} , \alpha_{\cal I} \ra = \delta_{\cal I}^{\cal J} &,& \quad \int \la \alpha_{\cal I} , \alpha_{\cal J} \ra = 0 = \int \la \beta^{\cal I} , \beta^{\cal J} \ra \,\,\,\, , \nn \\
\int \la \omega_{\cal A} , \tilde{\omega}^{\cal B} \ra = - \int \la \tilde{\omega}^{\cal B} , \omega_{\cal A} \ra = \delta_{\cal A}^{\cal B} &,& \quad \int \la \omega_{\cal A} , \omega_{\cal B} \ra = 0 = \int \la \tilde{\omega}^{\cal A} , \tilde{\omega}^{\cal B} \ra \,\,\,\, ,
\eea
where $\la \,, \ra$ is the Mukhai pairing defined by $\la \varphi , \psi \ra = - \varphi_1 \wedge \psi_5 + \varphi_3 \wedge \psi_3 - \varphi_5 \wedge \psi_1$ for odd forms and by $\la \varphi , \psi \ra = \varphi_0 \wedge \psi_6 - \varphi_2 \wedge \psi_4 + \varphi_4 \wedge \psi_2 - \varphi_6 \wedge \psi_0$ for even forms with $\varphi_p$ being the $p$-form component of the mixed-degree form $\varphi$ and similarly for $\psi$. 

For our purposes, it is very important that both the K\"{a}hler and complex structure moduli spaces are special K\"{a}hler manifolds. The relevant K\"{a}hler potentials are \cite{GLW}:
\bea
K_J &=& - \ln i \int \la e^{-J_c} , e^{-\bar{J}_c} \ra = - \ln i \left( \bar{X}^{\cal A} G_{\cal A} - X^{\cal A} \bar{G}_{\cal A} \right) , \nn \\
K_{\Omega} &=& - \ln i \int \la \Omega , \bar{\Omega} \ra = - \ln i \left( \bar{X}^{\cal I} G_{\cal I} - X^{\cal I} \bar{G}_{\cal I} \right) .
\eea
It is also important that now the basis forms are not closed, unlike in the CY case. Instead, we have \cite{GLW}:
\be \label{dbasis}
d \omega_{\alpha} = m_{\alpha}^{\cal I} \alpha_{\cal I} - e_{{\cal I} \alpha} \beta^{\cal I} \, , \qquad d \tilde{\omega}^{\alpha} = 0 \, , \qquad d\alpha_{\cal I} = e_{{\cal I} \alpha} \tilde{\omega}^{\alpha} \, , \qquad d\beta^{\cal I} = m^{\cal I}_{\alpha} \tilde{\omega}^{\alpha} \, ,
\ee
where we have used the notation $\omega_{\cal A} = (1 , \omega_{\alpha})$ and $\tilde{\omega}^{\cal A} = (\tilde{\omega}^{\alpha} , \star 1)$ with $\omega_{\alpha}$ $(\tilde{\omega}^{\alpha})$ being a basis for the 2- (4-) forms. Finally, in (\ref{dbasis}) $m_{\alpha}^{\cal I}$ and $e_{\alpha {\cal I}}$ are constant matrices satisfying $m_{\alpha}^{\cal I} e_{{\cal I} \beta} - e_{{\cal I} \alpha} m^{\cal I}_{\beta} = 0$. 

Now, the effective 4d superpotential of a heterotic compactification on an $SU(3)$ structure manifold has the form $W = \int (H + dJ_c)\wedge \Omega$, where $H$ is the NS flux. One can write this as:
\be
W = m_{\cal A}^{\cal I} G_{\cal I} (z) X^{\cal A} (t) - e_{\cal I A} X^{\cal I} (z) X^{\cal A} (t) \, ,
\ee 
where $m_0^{\cal I}$, $e_{{\cal J} 0}$ arise from the expansion of $H$ in the $(\alpha_{\cal I} , \beta^{\cal J})$ basis and $X^0 (t) = 1$\,, \,$X^{\alpha} (t) = t^{\alpha}$ (for details, see \cite{LA}). Denoting $L^{\cal I} = e^{K_{\Omega}/2} X^{\cal I}$\,, \,$M_{\cal I} = e^{K_{\Omega}/2} G_{\cal I}$ and $L^{\cal A} =e^{K_J/2} X^{\cal A}$\,, \,$M_{\cal A} = e^{K_J/2} G_{\cal A}$, the function $Z = e^{K/2} W$ (called "central charge" in analogy to the BH case) acquires the form:
\be
Z = e^{(K_{\Omega} + K_J)/2} W = m_{\cal A}^{\cal I} M_{\cal I} L^{\cal A} - e_{\cal I A} L^{\cal I} L^{\cal A} \, .
\ee
In terms of this function and the double-symplectic section ${\hat{\cal V}} = e^{(K_{\Omega} + K_J)/2} (\Omega \otimes e^{-J_c})$, one can write the following attractor equations for the heterotic string:\footnote{This is slightly different from a previous proposal in \cite{GD}.}
\be \label{HetAttr}
\hat{{\cal Q}} = 2 {\rm Re} ( \bar{Z} \hat{{\cal V}} + g^{i \bar{j}} g^{\alpha \bar{\beta}} D_i D_{\alpha} \hat{{\cal V}} \bar{D}_{\bar{j}} \bar{D}_{\bar{\beta}} \bar{Z} ) \, ,
\ee
where $\hat{{\cal Q}} = - m_{\cal A}^{\cal I} \alpha_{\cal I} \otimes \tilde{\omega}^{\cal A} + e_{\cal I A} \beta^{\cal I} \otimes \tilde{\omega}^{\cal A}$. In \cite{LA}, it was verified that (\ref{HetAttr}) implies automatically the susy conditions $D_i Z = 0$ and $D_{\alpha} Z = 0$. Furthermore, (\ref{HetAttr}) also implies that $Z=0$ \cite{LA}, i.e. Minkowski vacua. Hence (\ref{HetAttr}) encodes all supersymmetric flux vacua of the heterotic string at the classical level. (Recall that these vacua are necessarily Minkowski \cite{AS}; to have susy AdS vacua, one needs to include quantum effects like, for example, gaugino condensation \cite{FLi}.)

\section{Type II on $SU(3)\times SU(3)$ structure}
\setcounter{equation}{0}

Let us now consider type IIA/B strings compactified on spaces with $SU(3)\times SU(3)$ structure. As already mentioned above, the internal geometry in this case is characterized by a pair of pure spinors $\Phi_+$, $\Phi_-$ that generalizes the pair $J$, $\Omega$ of the previous section. The special case of $SU(3)$ structure is recovered by taking a diagonal $SU(3)$ subgroup; in this case $\Phi_-$ reduces to $\Omega$ and $\Phi_+$ reduces to $e^{-J_c}$. In general, however, $\Phi_-$ is a sum of 1-, 3- and 5-forms just like $\Phi_+$ is a sum of 0-, 2-, 4- and 6-forms.

Similarly to (\ref{JOexp}), we have the expansions:
\be
\Phi_+ = X^{\cal A} (t) \omega_{\cal A} - G_{\cal A} (t) \tilde{\omega}^{\cal A} \, , \qquad \Phi_- = X^{\cal I} (z) \alpha_{\cal I} - G_{\cal I} (z) \beta^{\cal I} \, ,
\ee
where, as in the previous section, the basis forms satisfy relations (\ref{pairings}), but with appropriately extended range for the ${\cal I}$ indices in order to encompass the basis for 1-, 3- and 5-forms. The $\Phi_{\pm}$ moduli spaces are special K\"{a}hler with K\"{a}hler potentials \cite{GLW}:
\be
K_+ (t) = - \ln i \int \la \Phi_+ , \bar{\Phi}_+ \ra \qquad {\rm and} \qquad K_- (z) = - \ln i \int \la \Phi_- , \bar{\Phi}_- \ra \, .
\ee
The analogue of (\ref{dbasis}) is now given by \cite{GLW}:
\bea \label{bigD}
&&{\cal D} \omega_{\cal A} \sim m_{\cal A}^{\cal I} \alpha_{\cal I} - e_{\cal I A} \beta^{\cal I} \, , \qquad \hspace*{0.15cm}{\cal D} \tilde{\omega}^{\cal A} \sim \tilde{m}^{\cal I A} \alpha_{\cal I} - \tilde{e}^{\cal A}_{\cal I} \beta^{\cal I} \, , \nn \\
&&{\cal D} \alpha_{\cal I} \sim -\tilde{e}_{\cal I}^{\cal A} \omega_{\cal A} + e_{\cal I A} \tilde{\omega}^{\cal A} \, , \qquad {\cal D} \beta^{\cal I} \sim - \tilde{m}^{\cal I A} \omega_{\cal A} + m^{\cal I}_{\cal A} \tilde{\omega}^{\cal A} \, ,
\eea
where "$\sim$" means equality up to terms that vanish under the symplectic pairing (\ref{pairings}) and ${\cal D}$ is an extension of the exterior differential that is due to nonzero NS flux and/or non-geometricity of the background\footnote{Non-geometric backgrounds differ from the geometric ones in that their transition functions contain string dualities, like T-duality. Note that non-geometric backgrounds are, in fact, necessary in order to have {\it all} charge components in (\ref{bigD}) non-vanishing.}; the constant charge matrices $e_{\cal I A}$, $m^{\cal I}_{\cal A}$, $\tilde{e}^{\cal A}_{\cal I}$ and $\tilde{m}^{\cal I A}$ satisfy appropriate constraints so that ${\cal D}^2 = 0$. For more details on these constraints and the precise form of ${\cal D}$, see \cite{LA} and references therein. 

Generically, type II on $SU(3)\times SU(3)$ structure gives an $N=2$ effective theory. One can obtain an $N=1$ truncation by considering orientifolds of these generalized compactifications \cite{BG}. The $N=1$ truncation can also be due to a spontaneous partial susy breaking \cite{CB}. Regardless of the mechanism, one can derive a compact Gukov-Vafa-Witten type formula for the superpotential of the effective 4d $N=1$ theory. Let us for concreteness focus on type IIA from now on. (We will comment on type IIB at the end.) Then \cite{CB}: 
\be \label{Z0}
e^{K/2} W = c \,e^{\frac{K_+}{2}+2\varphi} \int \la \Phi_+ , {\cal D} \Pi_- + G^{fl} \ra \, ,
\ee
where $c$ is a constant, $\varphi$ is the 4d dilaton, $G^{fl}$ is a sum of all internal RR fluxes (rescaled by a factor of $\sqrt{2}$ compared to \cite{CB, LA} for convenience) and the object $\Pi_-$ is
\be
\Pi_- = A^{odd}_{RR} + i {\rm Im} (C \Phi_-)
\ee
with $A^{odd}_{RR}$ being the sum of all internal RR potentials (again, rescaled by a factor of $\sqrt{2}$ compared to \cite{CB, LA}) and $C=const\times e^{-\phi}$ with $\phi$ being the 10d dilaton. Clearly, the proper $N=1$ variables arise from the expansions of $\Phi_+$ and $\Pi_-$, instead of $\Phi_+$ and $\Phi_-$. This makes things significantly more complicated as the space of deformations of $\Pi_-$ is {\it not} special K\"{a}hler. Nevertheless, it is K\"{a}hler with a K\"{a}hler potential given by \cite{GL, BG}:\footnote{Note that this corrects an error in \cite{LA}, where in the vein of \cite{GD} it was stated that $\hat{K}_- = -2 \ln i \int \la \Pi_- , \bar{\Pi}_- \ra$. \label{ftnt}}
\be \label{Kh}
\hat{K}_- = -2 \ln i \int \la C\Phi_-, \overline{C \Phi}_- \ra = 4 \varphi \, .
\ee
The last expression is a rather involved function of the $N=1$ K\"{a}hler coordinates $\{ \hat{X}^{\cal I} , \hat{G}_{\cal J} \}$, that are defined via the expansion $\Pi_- = \hat{X}^{\cal I} \alpha_{\cal I} - \hat{G}_{\cal I} \beta^{\cal I}$. Actually, at first sight it may not at all be obvious that (\ref{Kh}) depends on the correct variables. To see that it does, note that only half of ${\rm Re} (C \Phi_-)$ and ${\rm Im} (C \Phi_-)$ should be viewed as independent because of the way the Hodge star acts on $\Phi_-$. In particular, we can view ${\rm Re} (C \Phi_-)$ as functions of ${\rm Im} (C \Phi_-)$. To make this more clear, let us take the simplest example of $\Phi_-$, which is the holomorphic 3-form $\Omega$ of a CY(3) manifold. Now, due to $* \Omega = - i \Omega$, one has that \,$* {\rm Re} \Omega = {\rm Im} \Omega$ \,and \,$* {\rm Im} \Omega = - {\rm Re} \Omega$\,. Hence, in particular, we can view ${\rm Re} \Omega$ as determined by \,${\rm Im} \Omega$ \,via \,${\rm Re} \Omega = f ({\rm Im} \Omega) = - * {\rm Im} \Omega$. For more details on the general argument for a pure spinor, see \cite{NH, CB}. Therefore, $\hat{K}_-$ in (\ref{Kh}) should be viewed as a function of the variables in the expansion of ${\rm Im} C \Phi_-$, which is also ${\rm Im} \Pi_-$. That there is no dependence on ${\rm Re} \Pi_-$, which comes from the RR potentials, is in complete analogy with the fact that $K_J = - \ln i \int \la e^{-J_c}, e^{-\bar{J}_c}\ra = - \ln \frac{4}{3} \int J\wedge J\wedge J$ does not depend on the NS B-field moduli. This, clearly, means that the moduli space directions originating from the ${\rm Re} \Pi_-$ expansion correspond to shift symmetries of the metric determined by $\hat{K}_-$ \cite{BG, GL}.

Now, let us first consider the case of vanishing RR flux. Then, introducing $\hat{L}^{\cal I} = e^{\hat{K}_-/2} \hat{X}^{\cal I}$\,, \,$\hat{M}_{\cal I} = e^{\hat{K}_-/2} \hat{G}_{\cal I}$ and $L^{\cal A} =e^{K_+/2} X^{\cal A}$\,, \,$M_{\cal A} = e^{K_+/2} G_{\cal A}$, we can write (\ref{Z0}) as:
\be \label{Z}
Z= e^{K/2} W = c \left( \hat{L}^{\cal I} e_{\cal IA} L^{\cal A} - \hat{M}_{\cal I} m^{\cal I}_{\cal A} L^{\cal A} - \hat{L}^{\cal I} \tilde{e}_{\cal I}^{\cal A} M_{\cal A} + \hat{M}_{\cal I} \tilde{m}^{\cal IA} M_{\cal A} \right) .
\ee
In terms of this "central charge" and the appropriate analogue, ${\cal U} = e^{(\hat{K}_- + K_+)/2} (\Pi_- \oplus \Phi_+)$, of the heterotic double-symplectic section $\hat{{\cal V}}$, the attractor equations for the present case are:
\be \label{IIattr}
\hat{\cal Q} = \frac{2}{c} N^{-1} {\rm Re} (\bar{Z} {\cal U} + g^{\hat{i} \bar{\hat{j}}} g^{\alpha \bar{\beta}} D_{\hat{i}} D_{\alpha} {\cal U} \, \bar{D}_{\bar{\hat{j}}} \bar{D}_{\bar{\beta}} \bar{Z}) \, ,
\ee
where $\hat{{\cal Q}} = \tilde{m}^{\cal I A} \alpha_{\cal I} \otimes \omega_{\cal A} - \tilde{e}^{\cal A}_{\cal I} \beta^{\cal I} \otimes \omega_{\cal A} - m_{\cal A}^{\cal I} \alpha_{\cal I} \otimes \tilde{\omega}^{\cal A} + e_{\cal I A} \beta^{\cal I} \otimes \tilde{\omega}^{\cal A}$, the index $\hat{i}$ runs over the set of independent variables $\{ \hat{X}^{\cal I}, \hat{G}_{\cal I} \}$ and $N$ is the normalization of ${\cal U}$. Note that, unlike for the BH and heterotic attractors, this normalization is not a constant. More precisely, we have:
\be \label{norm}
N = \int \la \bar{\cal U} , {\cal U} \ra = - i e^{\hat{K}_-} \int \la \Pi_- , \bar{\Pi}_- \ra = - \frac{ \int || \Pi_- ||^2 vol_6 }{ \left( \int || C \Phi_- ||^2 vol_6 \right)^2 } \,\, ,
\ee 
which generically is a function of all of the variables $\{\hat{X}^{\cal I}, \bar{\hat{X}}^{\cal I}, \hat{G}_{\cal I}, \bar{\hat{G}}_{\cal I}\}$; in the last equality in (\ref{norm}), we have used (\ref{Kh}).\footnote{We should note that in \cite{LA} the function $N$ was equal to $-e^{\hat{K}_-/2}$ because of the error in the form of $\hat{K}_-$, which we mentioned in footnote \ref{ftnt}.} Let us also mention that (\ref{Z}) can be written as $Z = c \int \la \hat{{\cal Q}}, {\cal U} \ra$.

The attractor equations (\ref{IIattr}) can be shown to imply the susy conditions $D_{\alpha} Z = 0$, $D_{\hat{X}^{\cal I}} Z = 0$ and $D_{\hat{G}_{\cal I}} Z = 0$ for Minkowski vacua {\it only}. Namely, the condition $D_{\alpha} Z = 0$ is due to the special K\"{a}hler geometry of the $\Phi_+$ moduli space regardless of the value of $Z$ \cite{LA}. However, the other two conditions are only satisfied upon setting $Z=0$. Indeed, if we substitute the expressions for the charges from (\ref{IIattr}) into $D_{\hat{X}^{\cal I}} Z$, computed from (\ref{Z}), we find:
\be
D_{\hat{X}^{\cal I}} Z = c \left( e_{\cal I A} L^{\cal A} - \tilde{e}_{\cal I}^{\cal A} M_{\cal A} \right) e^{\frac{\hat{K}_-}{2}} + (\pd_{\hat{X}^{\cal I}} \hat{K}_-) Z = \left( i N^{-1} \bar{\hat{M}}_{\cal I} \,e^{\frac{\hat{K}_-}{2}} + (\pd_{\hat{X}^{\cal I}} \hat{K}_-) \right) \!Z \, ,
\ee
which is generically nonzero. (Note that $\pd_{\hat{X}^{\cal I}} \hat{K}_- = -4i e^{\frac{\hat{K}_-}{2}} {\rm Im} (C G_{\cal I})$ \cite{GL}.) So to ensure $D_{\hat{X}^{\cal I}} Z = 0$, we have to take $Z=0$. Similarly, the condition $D_{\hat{G}_{\cal I}} Z = 0$ is also satisfied only for $Z=0$. Therefore, the attractors (\ref{IIattr}) encode only Minkowski vacua. There is, however, a very important difference with the heterotic case. Namely, the heterotic attractors (\ref{HetAttr}) arise from the most general expansion of $\hat{{\cal Q}}$ in the basis of $\hat{{\cal V}}$, $D_i \hat{{\cal V}}$, $D_{\alpha} \hat{{\cal V}}$, $D_i D_{\alpha} \hat{{\cal V}}$ upon substituting the susy conditions $D_i Z = 0$ and $D_{\alpha} Z = 0$. On the other hand, the type II attractors (\ref{IIattr}) {\it do not} originate from the most general expansion. Indeed, since the moduli space of $\Pi_-$ is {\it not} special K\"{a}hler, the general expansion could contain terms proportional to $D_{\hat{i}} D_{\hat{j}} {\cal U}$, \,$D_{\hat{i}} D_{\hat{j}} D_{\hat{k}} {\cal U}$ \,etc. The structure of such terms and the question of whether their presence would allow for supersymmetric AdS vacua to be encoded deserve a thorough investigation. We hope to come back to these issues in the future. 

So far we have considered only vanishing RR flux. Let us now take $G^{fl} \neq 0$ in (\ref{Z0}) and focus on the contribution this leads to:
\be
Z^{RR} \equiv c \,e^{\frac{K_+}{2}+\frac{\hat{K}_-}{2}} \int \la \Phi_+ , G^{fl} \ra \, .
\ee 
Since we are in the type IIA case, the RR flux is a sum of even forms only and therefore it can be expanded as $G^{fl} = m_{RR}^{\cal A} \omega_{\cal A} - e_{RR, {\cal A}} \tilde{\omega}^{\cal A}$, where $m_{RR}^{\cal A}$ and $e_{RR, {\cal A}}$ are RR charges. Hence $Z^{RR}$ becomes:
\be \label{ZRR}
Z^{RR} = c e^{\frac{\hat{K}_-}{2}} (M_{\cal A} m_{RR}^{\cal A} - L^{\cal A} e_{RR, {\cal A}}) \, .
\ee
It is then straightforward to show that the attractor equations
\be \label{RRch}
m_{RR}^{\cal A} = c^{-1} e^{-\frac{\hat{K}_-}{2}} (\bar{Z}^{RR} L^{\cal A} - Z^{RR} \bar{L}^{\cal A}) \, , \quad \,\, e_{RR, {\cal A}} = c^{-1} e^{-\frac{\hat{K}_-}{2}} (\bar{Z}^{RR} M_{\cal A} - Z^{RR} \bar{M}_{\cal A})
\ee
imply that $D_{\alpha} Z^{RR} = 0$. Indeed, from (\ref{ZRR}) we have $D_{\alpha} Z^{RR} = c e^{\frac{\hat{K}_-}{2}} (m_{RR}^{\cal A} D_{\alpha} M_{\cal A} - e_{RR, {\cal A}} D_{\alpha} L^{\cal A})$. The last expression, upon substituting the RR charges from (\ref{RRch}), can be easily seen to vanish due to the special K\"{a}hler geometry of the $\Phi_+$ moduli space. (More precisely, due to $\bar{L}^{\cal A} M_{\cal A} - L^{\cal A} \bar{M}_{\cal A} = -i$ and $L^{\cal A} D_{\alpha} M_{\cal A} - M_{\cal A} D_{\alpha} L^{\cal A} = 0$.) However, $D_{\hat{X}^{\cal I}} Z^{RR} = (\pd_{\hat{X}^{\cal I}} \hat{K}_-) Z^{RR}$ is nonzero unless $Z^{RR} = 0$, similarly to the case of vanishing RR flux. So, denoting $Z_{tot} = Z^g + Z^{RR}$ with $Z^g$ being the same as $Z$ in (\ref{Z}) (i.e., the geometric and NS flux contribution), we have that the susy condition $D_{\alpha} Z_{tot} = 0$ is implied by the attractor equations (\ref{IIattr}) and (\ref{RRch}). On the other hand, the susy condition $D_{\hat{X}^{\cal I}} Z_{tot} = 0$ is more involved. Namely, adding the contributions from $Z^g$ and $Z^{RR}$, we find:
\be
D_{\hat{X}^{\cal I}} Z_{tot} = i N^{-1} \bar{\hat{M}}_{\cal I} \,e^{\frac{\hat{K}_-}{2}} Z^g + (\pd_{\hat{X}^{\cal I}} \hat{K}_-) Z_{tot} \, .
\ee
Hence restricting to Minkowski vacua, i.e. taking $Z_{tot} = 0$, is not enough to ensure that the supersymmetry conditions are satisfied. One needs, in addition, that $Z^g = 0$ too. It would be interesting to explore the consequences of this constraint. And, of course, it is also worth investigating whether it is possible to satisfy \,$ i N^{-1} \bar{\hat{M}}_{\cal I} \,e^{\frac{\hat{K}_-}{2}} Z^g = - (\pd_{\hat{X}^{\cal I}} \hat{K}_-) Z_{tot}$ \,so that one would have susy AdS vacua. Let us also note that including the terms $D_{\hat{i}} D_{\hat{j}} {\cal U}$ \,etc., that we mentioned at the end of the previous paragraph, will of course modify the discussion of this paragraph as well. 

Finally, let us comment on the type IIB case. Now the superpotential and the relevant K\"{a}hler potentials are obtained from the IIA expressions above by the substitution $\Phi_+ \leftrightarrow \Phi_-$, together with the exchange of odd RR potentials (even RR fluxes) with even RR potentials (odd RR fluxes) \cite{GLW, BG}. Hence, the role of the IIA pair $\Phi_+$ and $\Pi_-$ is played in type IIB by the pair $\Phi_-$ and $\Pi_+ = A^{ev}_{RR} + i {\rm Im} (C \Phi_+)$. It is straightforward to repeat the considerations of the previous paragraphs for the present case. In particular, when the RR fluxes vanish one has the following attractor equations for Minkowski vacua:
\be
\hat{{\cal Q}} = \frac{2}{c} \tilde{N}^{-1} {\rm Re} (g^{i \bar{j}} g^{\hat{\alpha} \bar{\hat{\beta}}} D_i D_{\hat{\alpha}} \tilde{{\cal U}} \,\bar{D}_{\bar{j}} \bar{D}_{\bar{\hat{\beta}}} \bar{Z}) \, ,
\ee 
where
\be
\tilde{{\cal U}} = e^{(K_- + \hat{K}_+)/2} (\Phi_- \otimes \Pi_+) \, , \qquad \tilde{N} = \int \la \bar{\tilde{{\cal U}}} , \tilde{{\cal U}} \ra = - \frac{\int ||\Pi_+||^2 vol_6}{(\int || C \Phi_+||^2 vol_6)^2}
\ee
and the index $\hat{\alpha}$ runs over the set of independent $N=1$ K\"{a}hler coordinates $\{ \hat{X}^{\cal A} , \hat{G}_{\cal A} \}$ defined via the expansion $\Pi_+ = \hat{X}^{\cal A} \omega_{\cal A} - \hat{G}_{\cal A} \tilde{\omega}^{\cal A}$.

\section{Acknowledgements}
This paper is mostly based on a work presented by L.A. at the 4th RTN Workshop "Constituents, Fundamental Forces and Symmetries of the Universe" held in Varna, September 2008. The research of L.A. is supported by DOE grant FG02-84-ER40153.


\begin{thebibliography}{100}

\bibitem{DRS}
K. Dasgupta, G. Rajesh and S. Sethi, {\em M-theory, Orientifolds and G-flux}, JHEP {\bf 9908} (1999) 023, hep-th/9908088.

\bibitem{GKP}
S. Giddings, S. Kachru and J. Polchinski, {\em Hierarchies from Fluxes in String 
Compactifications}, Phys. Rev. {\bf D66} (2002) 106006, hep-th/0105097.

\bibitem{AS}
A. Strominger, {\em Superstrings with Torsion}, Nucl. Phys. {\bf B274} (1986) 253.

\bibitem{Het}
G. L. Cardoso, G. Curio, G. Dall'Agata, D. Lust, P. Manousselis and G. Zoupanos, {\em Non-K\"{a}hler String Backgrounds and their Five Torsion Classes}, Nucl. Phys. {\bf B652} (2003) 5, hep-th/0211118.

\bibitem{GLW}
M. Grana, J. Louis and D. Waldram, {\em Hitchin Functionals in $N=2$ Supergravity}, JHEP {\bf 0601} (2006) 008, hep-th/0505264; {\em $SU(3)\times SU(3)$ Compactification and Mirror Duals of Magnetic Fluxes}, JHEP {\bf 0704} (2007) 101, hep-th/0612237.

\bibitem{GMPT}
M. Grana, R. Minasian, M. Petrini and A. Tomasiello, {\em Generalized Structures of $N=1$ Vacua }, JHEP {\bf 0511} (2005) 020, hep-th/0505212.

\bibitem{NH}
N. Hitchin, {\em Generalized Calabi-Yau Manifolds}, Quart. J. Math. Oxford Ser. {\bf 54} (2003) 281, arXiv:math/0209099 [math.DG]; {\em The Geometry of Three-forms in Six and Seven Dimensions}, arXiv:math/0010054 [math.DG]; {\em Stable Forms and Special Metrics}, arXiv:math/0107101 [math.DG].

\bibitem{MG}
M. Gualtieri, {\em Generalized Complex Geometry}, arXiv:math/0401221 [math.DG].

\bibitem{CB}
D. Cassani and A. Bilal, {\em Effective Actions and $N=1$ Vacuum Conditions from $SU(3)\times SU(3)$ Compactifications}, JHEP {\bf 0709} (2007) 076, arXiv:0707.3125.

\bibitem{RK}
R. Kallosh, {\em New Attractors}, JHEP {\bf 0512} (2005) 022, hep-th/0510024.

\bibitem{LA}
L. Anguelova, {\em Flux Vacua Attractors and Generalized Compactifications}, JHEP {\bf 0901} (2009) 017 , arXiv:0806.3820 [hep-th].

\bibitem{BHattr}
S. Ferrara, R. Kallosh and A. Strominger, {\em $N=2$ Extremal Black Holes}, Phys. Rev. {\bf D52} (1995) 5412, hep-th/9508072; S. Ferrara and R. Kallosh, {\em Supersymmetry and Attractors}, Phys. Rev. {\bf D54} (1996) 1514, hep-th/9602136; A. Strominger, {\em Macroscopic Entropy of $N=2$ Extremal Black Holes}, Phys. Lett. {\bf B383} (1996) 39, hep-th/9602111.

\bibitem{GD}
G. Dall'Agata, {\em Non-K\"{a}hler Attracting Manifolds}, JHEP {\bf 0604} (2006) 001, hep-th/0602045.

\bibitem{FLi}
A. Frey and M. Lippert, {\em $AdS$ Strings with Torsion: Non-complex Heterotic Compactifications}, Phys. Rev. {\bf D72} (2005) 126001, hep-th/0507202; P. Manousselis, N. Prezas and G. Zoupanos {\em Supersymmetric Compactifications of Heterotic Strings with Fluxes and Condensates}, Nucl. Phys. {\bf B739} (2006) 85, hep-th/0511122.

\bibitem{BG}
I. Benmachiche and T. Grimm, {\em Generalized $N=1$ Orientifold Compactifications and the Hitchin Functionals}, Nucl. Phys. {\bf B748} (2006) 200, hep-th/0602241; 

\bibitem{GL}
T. Grimm and J. Louis, {\em The Effective Action of Type IIA Calabi-Yau Orientifolds}, Nucl. Phys. {\bf B718} (2005) 153, hep-th/0412277.

\end{thebibliography}
\end{document}